\documentclass{article}

\usepackage{amsmath}
\usepackage{graphicx}
\usepackage{cite}

\input epsf

\begin{document}

\setlength{\unitlength}{1mm}

\begin{titlepage}

\begin{flushright}
Edinburgh 2002/03\\
LAPTH--904/02\\
March 2002
\end{flushright}
\vspace{1.cm}

\begin{center}
\large\bf
{\LARGE \bf A next--to--leading order study of photon--pion and pion pair 
hadro--production in the light of the Higgs boson search at the LHC.}\\[2cm]
\rm
{ T.~Binoth$^{b}$, J.~Ph.~Guillet$^{a}$, E.~Pilon$^{a}$ and M. Werlen$^{a}$}
\\[.5cm]

{\em $^{a}$ Laboratoire d'Annecy-Le-Vieux de Physique Th\'eorique LAPTH,}\\
{\em Chemin de Bellevue, B.P. 110, F-74941  Annecy-le-Vieux, France}\\[.2cm]
{\em $^{b}$ Department of Physics and Astronomy,
University of Edinburgh, }\\
{\em  Edinburgh EH9 3JZ, Scotland, UK}        
      
\end{center}
\normalsize
\vspace{2cm}

\begin{abstract} 

We discuss the production of $\gamma \pi^{0}$ and $\pi^{0} \pi^{0}$ 
pairs with a 
large invariant mass at collider energies. We present a study based on a
perturbative QCD calculation at full next-to-leading order accuracy,
implemented in the computer programme {\it DIPHOX}. We give estimations for 
various observables, which concern the reducible background to the Higgs boson
search in the channel $H \to \gamma \gamma$, in the mass range 80--140 GeV at
the LHC. We critically discuss the reliability of these estimates due to 
our imperfect knowledge of fragmentation functions at high $z$ and a 
subtle interplay between higher order corrections 
and realistic experimental cuts. 
Whereas the invariant mass spectrum of photon--pion pairs 
is theoretically better under control, in the dipion case
large uncertainties remain.   
Finally we comment on the impact of our findings on Higgs boson searches
at the LHC. We conclude that the qualitative statement that the
pion backgrounds should not be dangerous for the $H\rightarrow \gamma\gamma$
search channel  remains true at the next--to--leading order level.   

\end{abstract}

\end{titlepage}

\section{Introduction}\label{intro}

A major motivation for the CERN Large
Hadron Collider (LHC) experiments\cite{atlas,cms,atlas+cms,cernreplhc} 
is to shed light
into the mechanism of electroweak symmetry breaking. 
As the LEP precision measurements favour a relatively light Higgs boson,
its decay into a pair of photons is a prominent search channel
which deserves intense consideration from the theoretical side.
The inclusion of QCD radiative corrections 
is mandatory not only for the signal 
but also for the background if one is aiming towards reliable predictions. 
In this respect it is important to note that especially at LHC, 
but also at the Tevatron, prompt photon signals --- photons which do
not stem from mesonic decays ---  are in general contaminated
by $\pi^0$ events.
The photon pairs from the mesonic decay which are very collimated,
if the pions have a high momentum compared to the pion mass, may appear
as single (fake) photons inside the detector. 
This background is especially worrisome in collider experiments, such as the Collider Detector at Fermilab (CDF) \cite{cdf} and D0 \cite{d0} experiments at the Fermilab Tevatron, where the pions cannot all be reconstructed event by event. At the LHC $\gamma \pi^{0}$ and $\pi^{0}
\pi^{0}$ production represent an important fraction of the reducible
background to the search of neutral Higgs bosons in the channel $H \rightarrow
\gamma \gamma$ in the intermediate mass range 80--140 GeV. Since the jet-jet cross section is about eight orders of magnitude larger than the expected Higgs boson signal for a standard model Higgs boson, this reducible background is overwhelming before any selection cut is imposed. 

\vspace{0.2cm}

\noindent
Regarding both issues of prompt diphoton production and Higgs boson search,
various cuts, especially isolation cuts in transverse energy have to be imposed
to the candidate events, in order to reject most of this background. Yet, the
actual fraction of this background which passes these cuts is not
negligible, and it is not well known. Up to now its estimations rely on Monte
Carlo simulations encoding the partonic subprocesses at the leading
logarithmic (LL) accuracy only. We may thus worry about large higher order
corrections which may be not accurately reproduced by
parton showers. Furthermore, the
hadronisation models matched with the partonic ingredients in these simulations
are tuned to describe correctly the bulk of pion production, but they may not
be accurate in the kinematical region relevant for the pions which pass the
cut. 
A tentatively quantitative study of isolated $\gamma \pi^{0}$ and $\pi^{0}
\pi^{0}$ production, confronted to a careful analysis of data collected at the
Tevatron could improve this situation. The motivation of this
article is to provide a study of full next-to-leading order (NLO) accuracy,
based on the computer code {\it DIPHOX} as an attempt in this direction. 

\vspace{0.2cm}

\noindent
The computer program {\it DIPHOX} on which our study relies is a Monte Carlo
code of partonic event generator type. This code has been designed to describe
the production of pairs of particles in hadronic collisions, accounting for all
contributing partonic processes at full NLO accuracy. These particles can be
prompt photons or hadrons, in particular neutral pions. This code is flexible
enough to accommodate various kinematic or calorimetric cuts. Especially, it
allows to compute cross sections for isolated $\gamma \pi^{0}$ and $\pi^{0}
\pi^{0}$  pairs, for any infrared  and collinear safe isolation criterion which
can be implemented at the partonic level. The physical content and schematical
description of {\it DIPHOX} has been given in Ref. \cite{bgpw},  where we have
concentrated on the case of diphoton production \cite{new14,houches99,new15}.  
In Ref. \cite{bgpw3} we recalled briefly how to use this code also in the
case of $\gamma \pi^{0}$ and $\pi^{0} \pi^{0}$ production, and we focused on
the phenomenology of dipion production at fixed target energies, with a
comparison to the data of the E706 experiment \cite{e706} at the Tevatron. 
A similar study was also presented in \cite{owens}.
The aim of the comparison was to confront the theoretical 
framework on which {\it
DIPHOX} relies to existing dipion data in order to validate this framework
before transporting  this knowledge to collider energies \cite{bgpw2} using the
factorisation property of perturbative QCD. 

\vspace{0.2cm}

\noindent
The main goal of the present article is to provide 
a NLO prediction for the reducible background to 
Higgs boson searches in the mass range
80--140 GeV at the LHC.   
In section 2 we present estimates for distributions which are of interest regarding the Higgs boson search, such as the distributions of  
invariant mass and transverse momentum of a pair. 
In section 3 we will extensively
discuss the uncertainties of our next--to--leading order approach
induced by severe experimental cuts. 
Finally section 4 gathers our conclusions and some outlook.

\section{Predictions for the $\gamma\pi^0$, $\pi^0\pi^0$ 
background at the LHC}\label{lhc}

In this section we discuss NLO predictions for the $\gamma \pi^{0}$ and $\pi^{0}
\pi^{0}$ reducible background to the neutral Higgs boson search 
channel $H \rightarrow \gamma \gamma$ in the intermediate mass range 80--140
GeV  at the LHC. All NLO curves where computed with the {\it DIPHOX} code,
version (1.2) 
\footnote{In this version a bug was removed compared to the first version 
described in \cite{bgpw}. In that paper the main issue 
was the prediction of photon pairs. Only  
fragmentation contributions in the case of severe isolation
cuts were affected by the bug. These are numerically irrelevant for the
diphoton rate predictions.}. 
 For all figures we are using the MRST2 \cite{MRST2} parton distribution functions, and, if not stated differently, the KKP fragmentation functions \cite{kkp}.  
For all plots given below the following cuts on the transverse
momenta, pseudo rapidities of the photons/pions and their invariant mass  
were applied: 
\begin{eqnarray}
p_T(\gamma_1,\pi_1) &>& 40 \quad \mbox{GeV}, \nonumber\\
p_T(\gamma_2,\pi_2) &>& 25 \quad \mbox{GeV}, \nonumber\\
|y(\gamma,\pi)| &<& 2.5, \nonumber\\
80 \; \mbox{GeV} &<& m_{\pi\pi,\gamma\pi,\gamma\gamma}\quad < \quad 140 \;\mbox{GeV}.
\end{eqnarray}
In addition, isolation cuts are imposed to reject events which
are accompanied by a certain amount of hadronic energy, as
this is typically not the case for photonic Higgs decay.
A photon is said to be
isolated if, in a cone in rapidity and azimuthal angle about the photon
direction, the amount of deposited hadronic transverse energy $E_{T}^{had}$ is
smaller than some  value $E_{T \, max}$ fixed by the experiment:
\begin{equation}\label{isol}   
E_{T}^{had} \leq E_{T \, max} \;\;\;\; \mbox{inside} \;\;\;\;    
\left( y - y_{\gamma} \right)^{2} +   
\left(  \phi - \phi_{\gamma} \right)^{2}  \leq R^{2}    
\end{equation}   
The actual experimental criteria, which are defined at the detector level, are
much more complicated than the simple one given by Eq. (\ref{isol}), see
\cite{cdf,d0}. They cannot be implemented in a partonic calculation. The
schematical criterion of Eq. (\ref{isol}) is a modelization of their effects
at the parton level. The topic of isolation of photons based on the above
criterion is extensively discussed in the literature 
\cite{boo1,abf,berger-qiu,gordon-vogelsang,vogt-vogelsang,cfgp}.  
We stress that there is a subtle interplay between higher order corrections
and the experimental cuts. These subtleties will be discussed
in the next section.

\vspace{0.2cm}

\begin{figure}[p]
\begin{center}
\includegraphics[width=\linewidth]{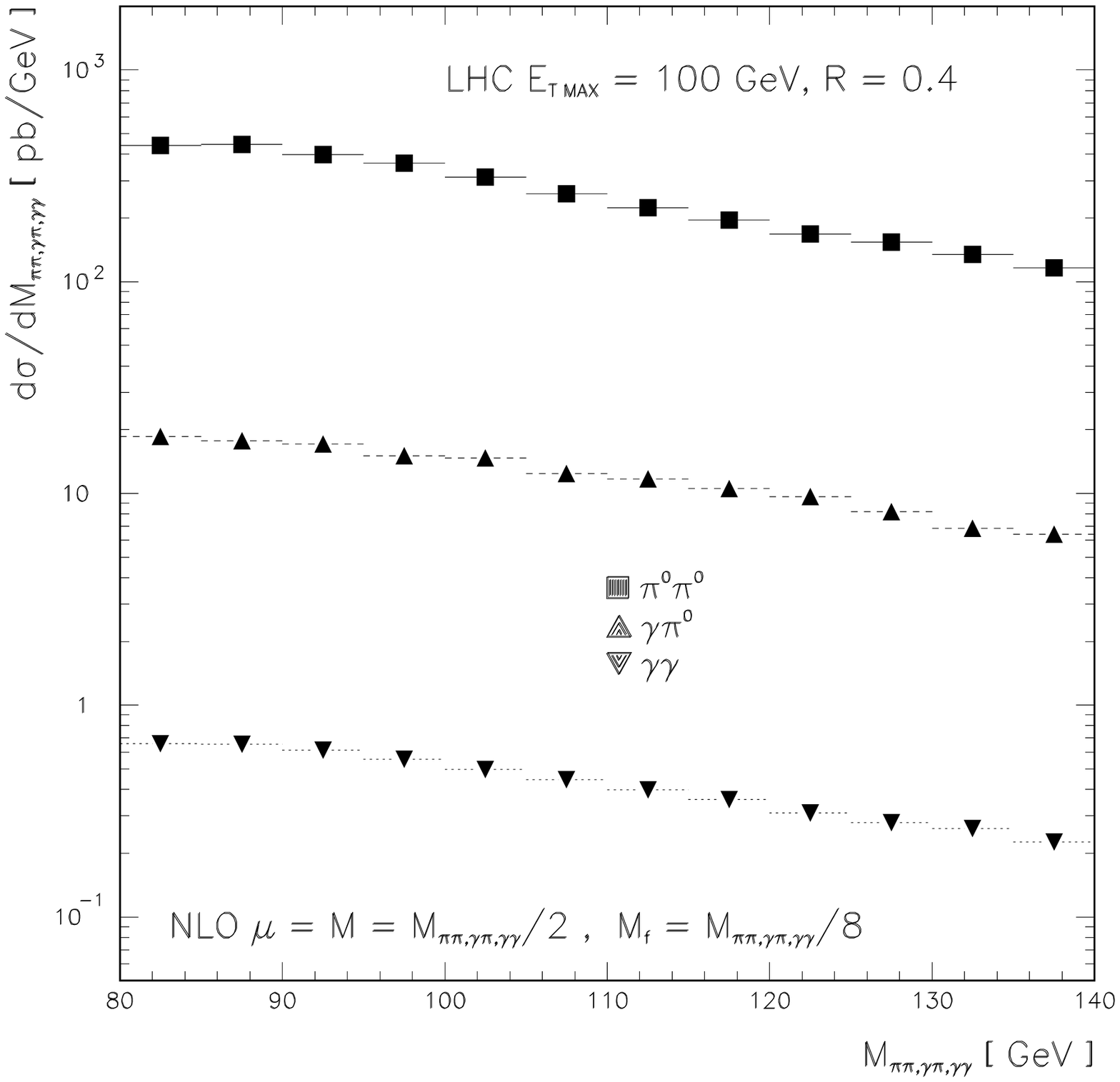}
\end{center}
\caption{\label{Fig:lhc_all_et00}{\em Comparison of the $\gamma\gamma$, $\gamma\pi^0$ and $\pi^0\pi^0$ invariant mass distribution for the 
isolation cuts $E_{T max} = 100$ GeV, R=0.4.}}
\end{figure}

\begin{figure}[p]
\begin{center}
\includegraphics[width=\linewidth]{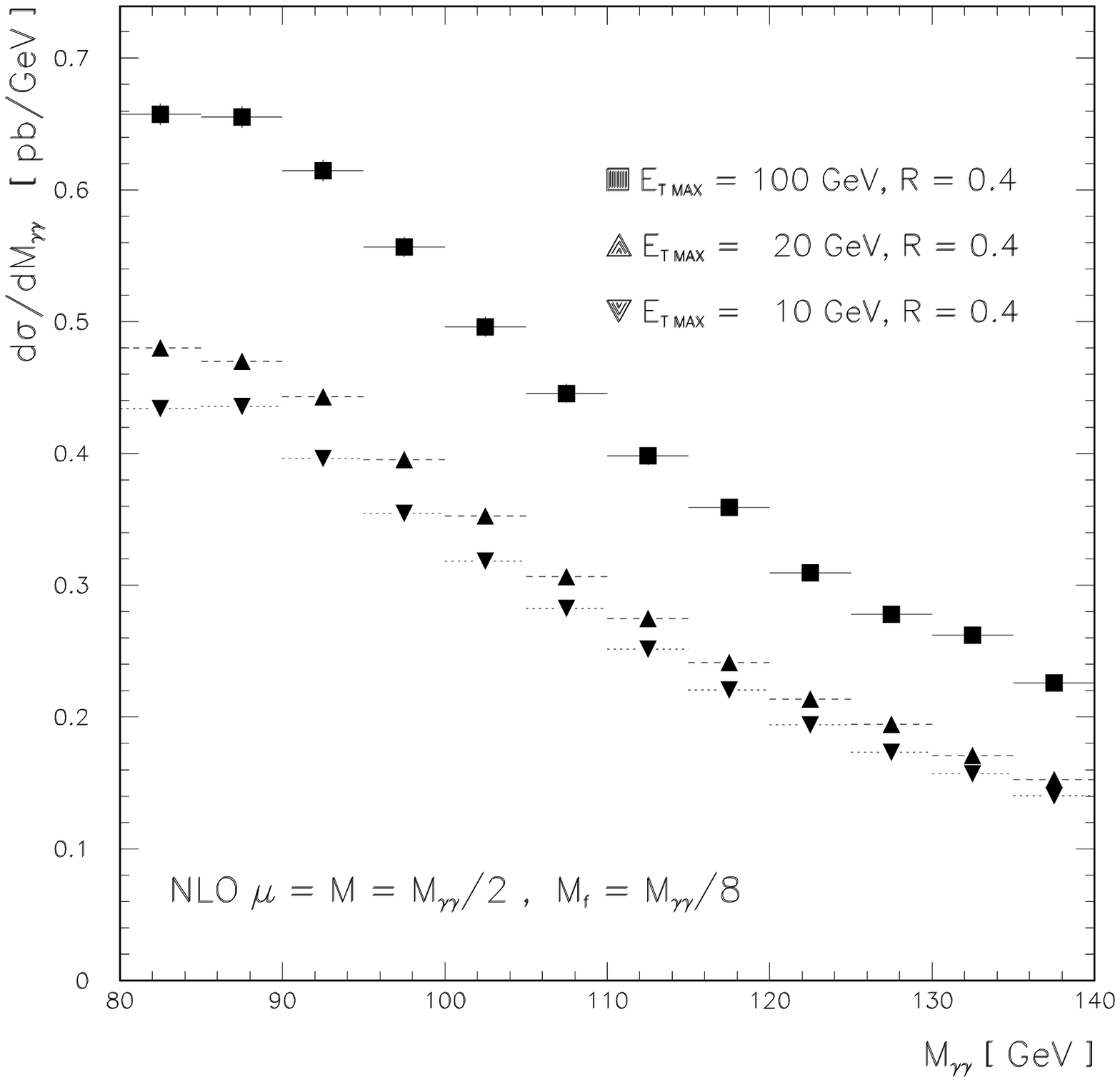}
\end{center}
\caption{\label{Fig:lhc_gg_mgg_isol}{\em Invariant mass distribution of prompt photon pairs for various values of $E_{Tmax}$.}}
\end{figure}

\begin{figure}[p]
\begin{center}
\includegraphics[width=\linewidth]{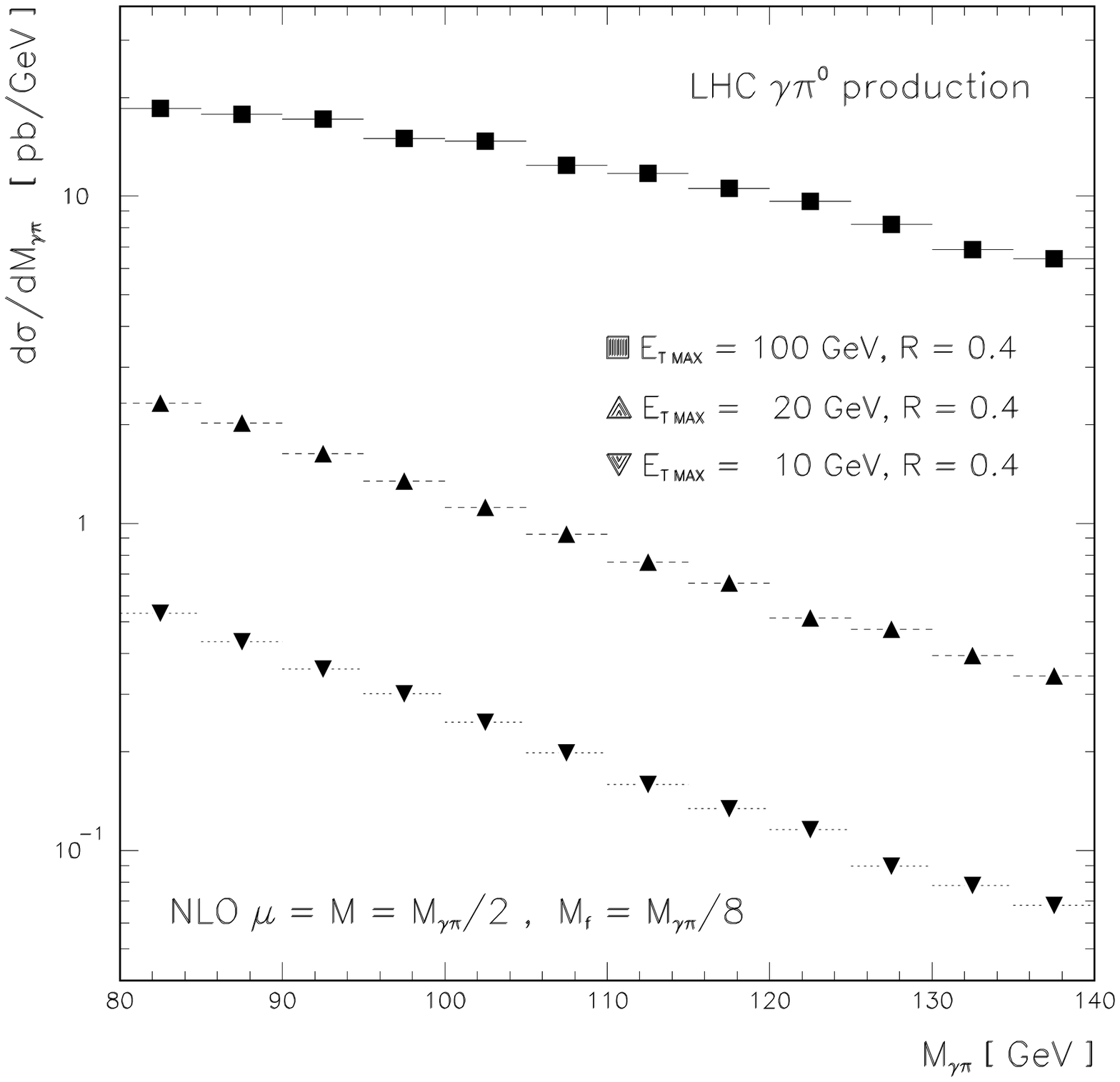}
\end{center}
\caption{\label{Fig:lhc_pg_mpg_isol}{\em Invariant mass distribution of  
photon--pion pairs for various values of $E_{Tmax}$.}}
\end{figure}

\begin{figure}[p]
\begin{center}
\includegraphics[width=\linewidth]{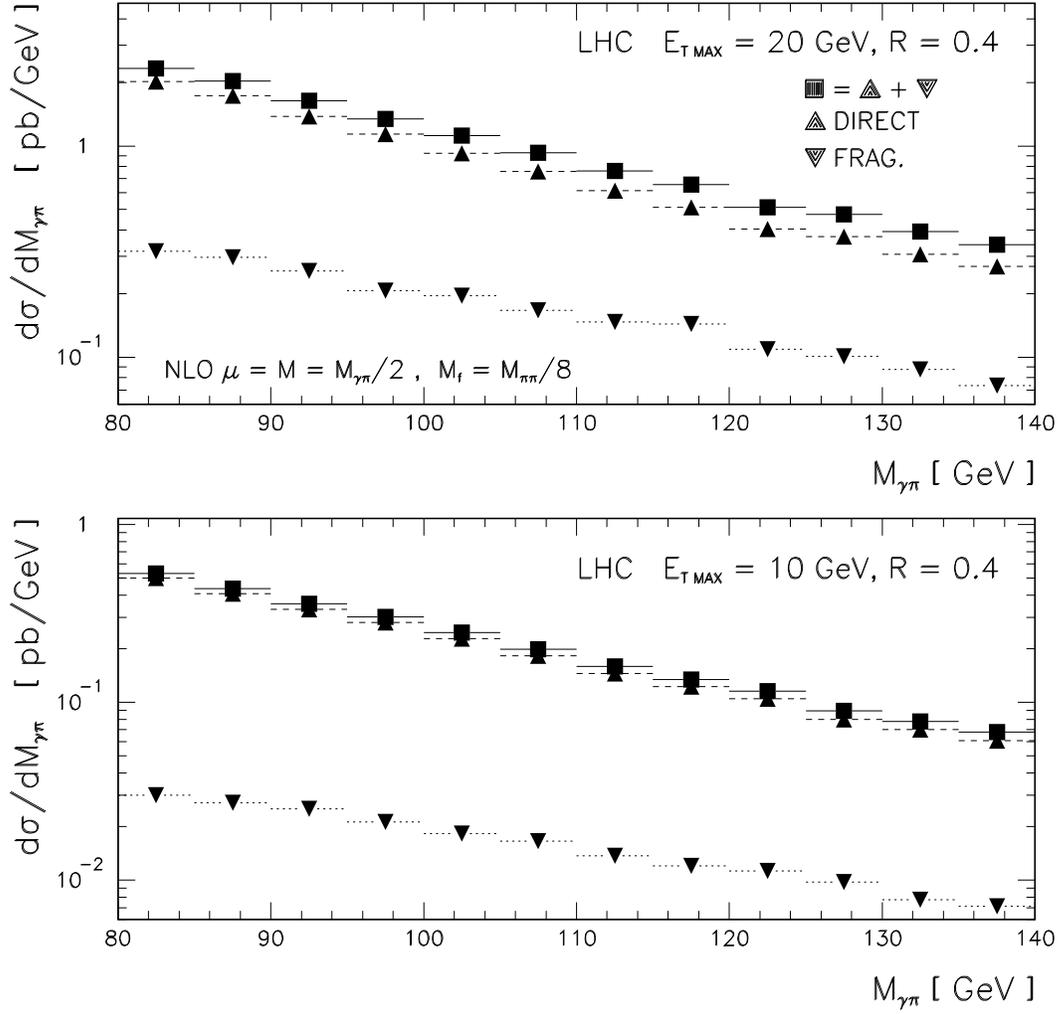}
\end{center}
\caption{\label{Fig:lhc_pg_mgp_dir_vs_fra}{\em The suppression of the fragmentation component of $\pi^0\gamma$ production through isolation.}}
\end{figure}

\begin{figure}[p]
\begin{center}
\includegraphics[width=\linewidth]{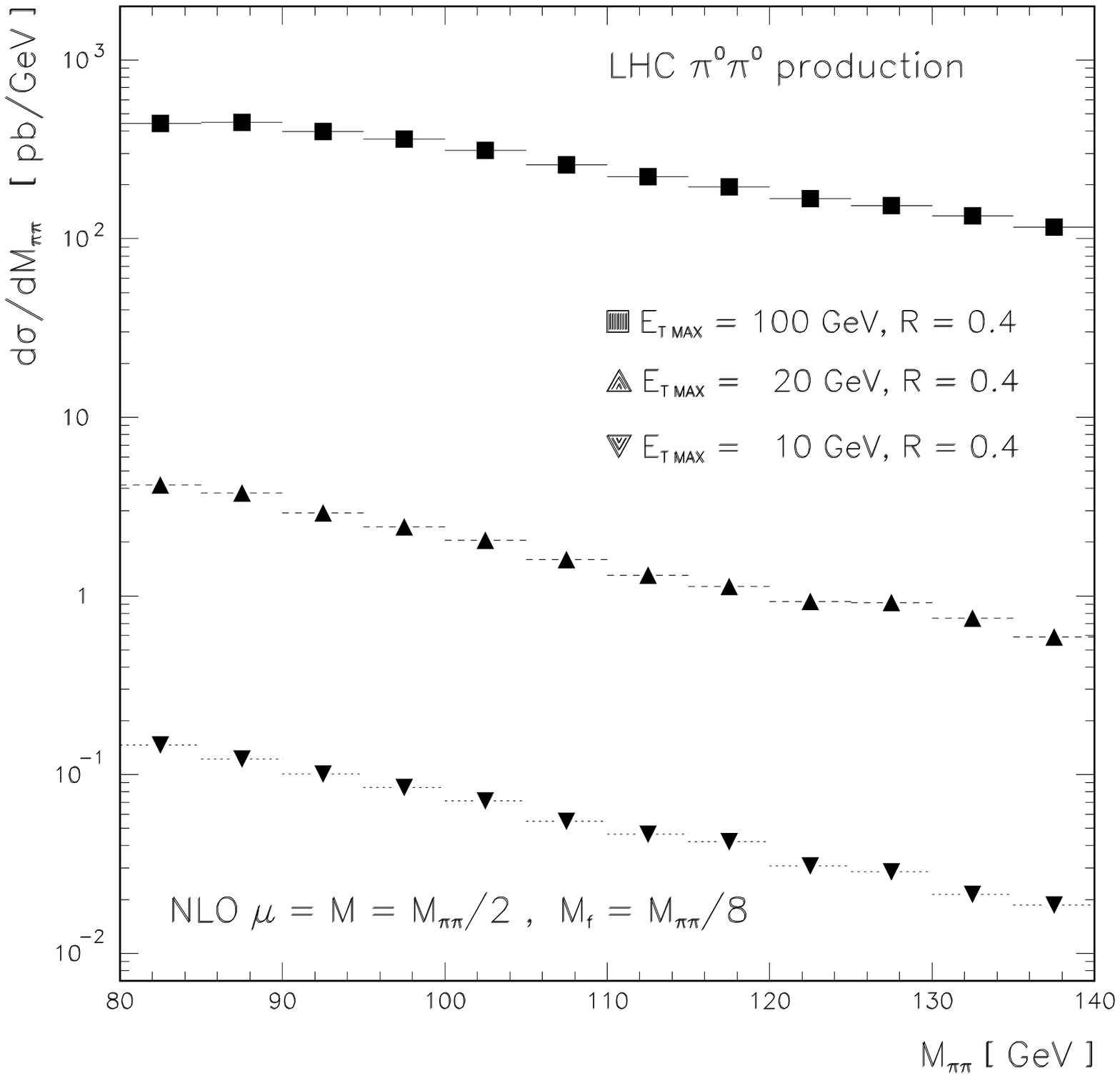}
\end{center}
\caption{\label{Fig:lhc_pp_mpp_isol}{\em Invariant mass distribution
of pion pairs for various values of $E_{Tmax}$.}}
\end{figure}

\noindent
In Fig.~\ref{Fig:lhc_all_et00} we show the invariant mass
spectra of $\gamma\gamma$, $\gamma\pi^0$ and $\pi^0\pi^0$ 
pairs for a very loose isolation criterion
of $E_{T\;max}=100$ GeV in a cone of $R=0.4$ at NLO. Note that the
fully inclusive case would be plagued by uncertainties stemming
from the low--$z$ regime of the pion fragmentation functions
(see discussion below). To avoid this source of uncertainty we
do not present figures showing the inclusive rate.
The plot shows the importance
of the reducible $\pi^0\gamma$, $\pi^0\pi^0$ backgrounds before isolation
cuts. The $\pi^0\gamma$ component is more than an order of magnitude 
above the $\gamma\gamma$ background. 
The $\pi^0\pi^0$ background is of the order of several hundreds of pb/GeV.
Tightening the isolation requirement by imposing $E_{T\;max}=10$ GeV 
changes the production rate of these components. This is shown in  Fig.~\ref{Fig:lhc_all_et10}.
We remind that the very narrow Higgs signal 
will be a sharp peak --- its width is determined by the detector
resolution of around 1 GeV --- ranging from 50 to 100 fb/GeV 
in the given mass window \cite{spira}.
The $\pi^0\pi^0$ background is reduced considerably whereas the 
$\gamma\pi^0$ background is still of the order
of the irreducible diphoton background. Concerning the prompt photon pairs,
the used isolation criterion reduces its rate by about a factor of two, because
the fragmentation component of the prompt photon pair is vetoed by isolation cuts. Note that for the given isolation criteria 
not only direct photon pairs remain. 
For  $E_{T max}=10$ GeV almost 20 per cent 
of the total rate still come from fragmentation in the given plot
for the scale choice used. 
Fig.~\ref{Fig:lhc_gg_mgg_isol} shows the suppression 
of the irreducible diphoton background when
isolation is imposed. 
To suppress photons from fragmentation more efficiently 
harder  isolation cuts than the ones shown in the plot  
have to be imposed.  A remaining uncertainty for the prompt 
photon pairs comes from higher order corrections to the box contribution. 
These corrections have been computed recently \cite{berndixonfreitas}
but are not yet implemented in {\it DIPHOX}. 
To give an idea of the rejection factors due to isolation 
we have plotted the 
invariant mass distributions of the photon--pion pair
for maximally allowed hadronic energies of $E_{T\;max}=100, 20$, and 10 GeV
in a cone of $R=0.4$ in Fig.~\ref{Fig:lhc_pg_mpg_isol}.  
Due to the collinear fragmentation model used the
hadronic energy is mainly located along the fragmenting
particle. This leads to the fact that there is no significant
$R$ dependence present. Only genuine higher order $2\rightarrow 3$
matrix elements can induce such a dependence and we only state that
they are on the per cent level. 
In the case of photon--pion production two mechanisms 
for the production of the photon are possible. 
The photon can be produced directly in a hard interaction
or through fragmentation. In the case of isolation the
latter component is  reduced by an order of magnitude
as can be seen in Fig.~\ref{Fig:lhc_pg_mgp_dir_vs_fra}. 
For completeness we provide with the plot for the invariant mass
distribution of pion pairs, Fig.~\ref{Fig:lhc_pp_mpp_isol}. 
In a realistic experimental situation 
the pion rates can be suppressed further.
In the relevant $p_T$ range, an extra reduction factor of 2 to 3 per pion
can be expected in fine-grained detectors with recognition of isolated $\pi^0$
as two overlapping showers  \cite{atlcal,cmsecal}.

\vspace{2mm}

\begin{figure}[p]
\begin{center}
\includegraphics[width=\linewidth]{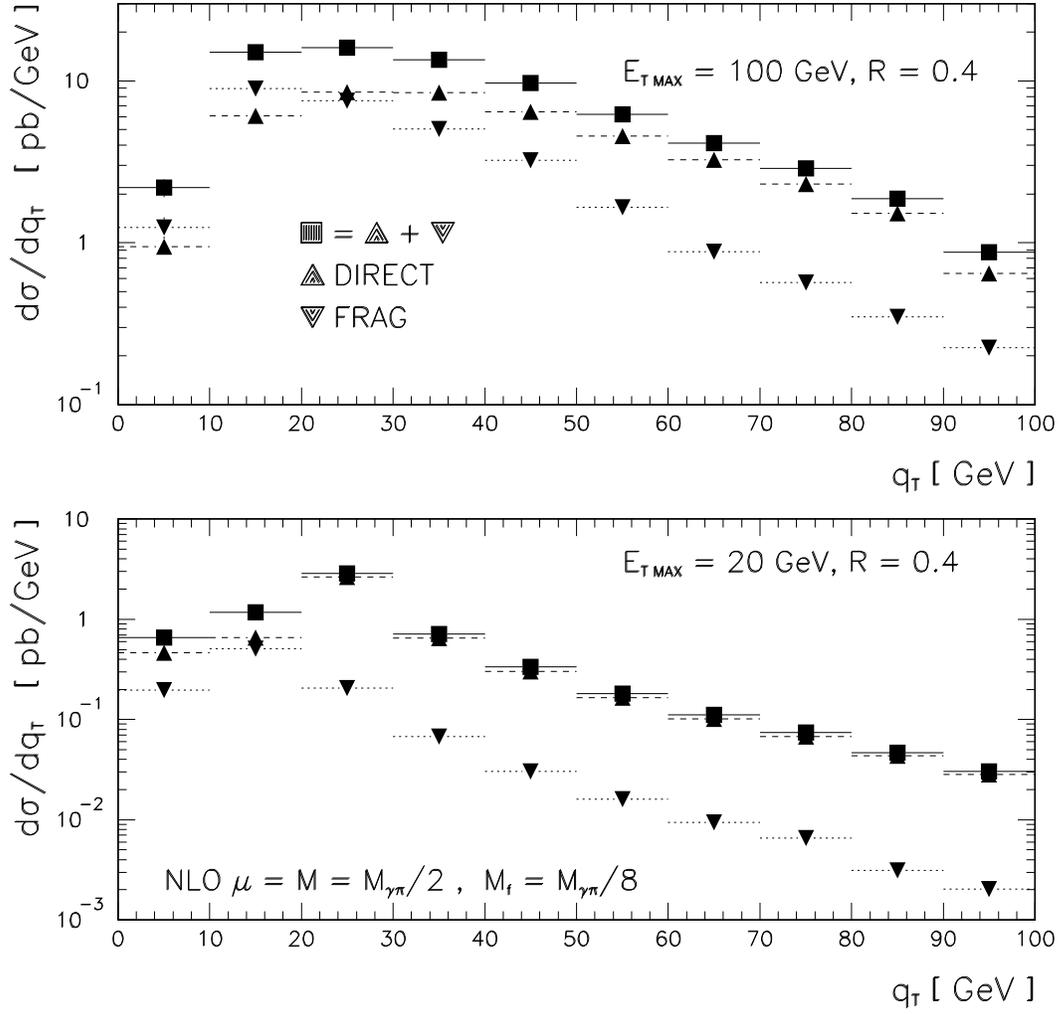}
\end{center}
\caption{\label{Fig:lhc_pg_qt_do}{\em The $q_{T}$ dependence of 
photon--pion pairs split into direct and fragmentation part with (top) loose isolation and (bottom) severe isolation.}}
\end{figure}

\begin{figure}[p]
\begin{center}
\includegraphics[width=\linewidth]{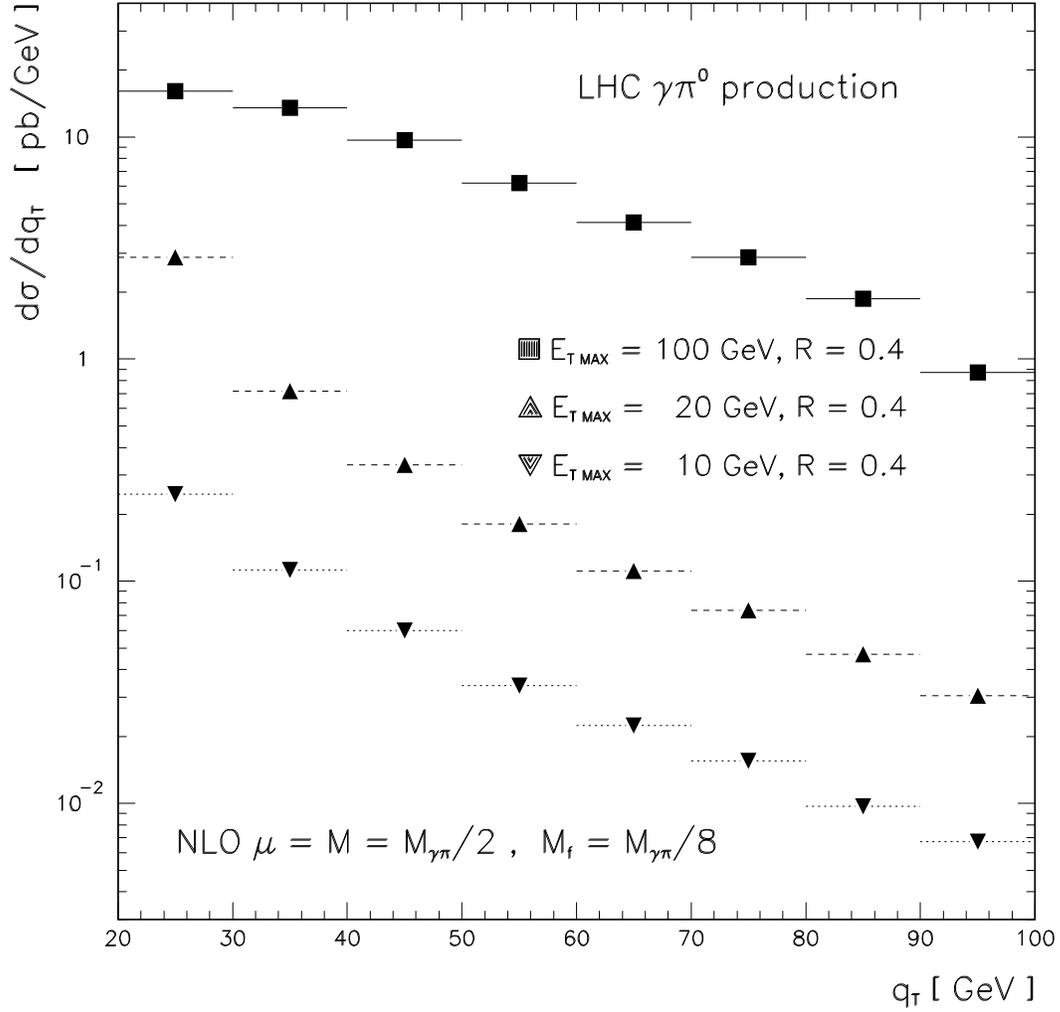}
\end{center}
\caption{\label{Fig:lhc_pg_qt_isol}{\em The $q_{T}$ dependence of 
photon--pion pairs for various isolation criteria.}}
\end{figure}

\noindent
Another spectrum of relevance is the transverse momentum
distribution of the pair produced particles,
$q_T=|\vec p_{T\;1}+\vec p_{T\;2}|$. In
Fig.~\ref{Fig:lhc_pg_qt_do} 
this distribution is shown for two isolation
criteria. 
Recall that the full pion photon rate is built up out of the 
direct component and the fragmentation component.
At colliders which operate at lower energies the fragmentation
component is typically only a fraction of the full rate.
Through isolation the fragmentation component dies out quickly.
Fig.~\ref{Fig:lhc_pg_qt_isol} gives an idea of
the suppression factors due to isolation. 
Note that this theoretical curve
is not reliable in the first few bins, as multiple soft
gluon emission alters the divergent behaviour of the fixed
order calculation. Near $q_T=0$ the partonic cross section
diverges logarithmically. This divergence is diluted by convoluting 
the partonic cross section 
with the pion fragmentation function. Also
at $q_T=E_{T\;max}$ there is a critical point inside the physical spectrum 
(see \cite{bgpw} for a discussion). The large bin width
prevents from observing this singular behaviour.  
To deal with these issues in a rigorous way is beyond the 
scope of this study\footnote{The corresponding
resummations to be done are technically very involved and no satisfying
procedure is defined yet.}. 

\vspace{2mm}

\begin{figure}[p]
\begin{center}
\includegraphics[width=\linewidth]{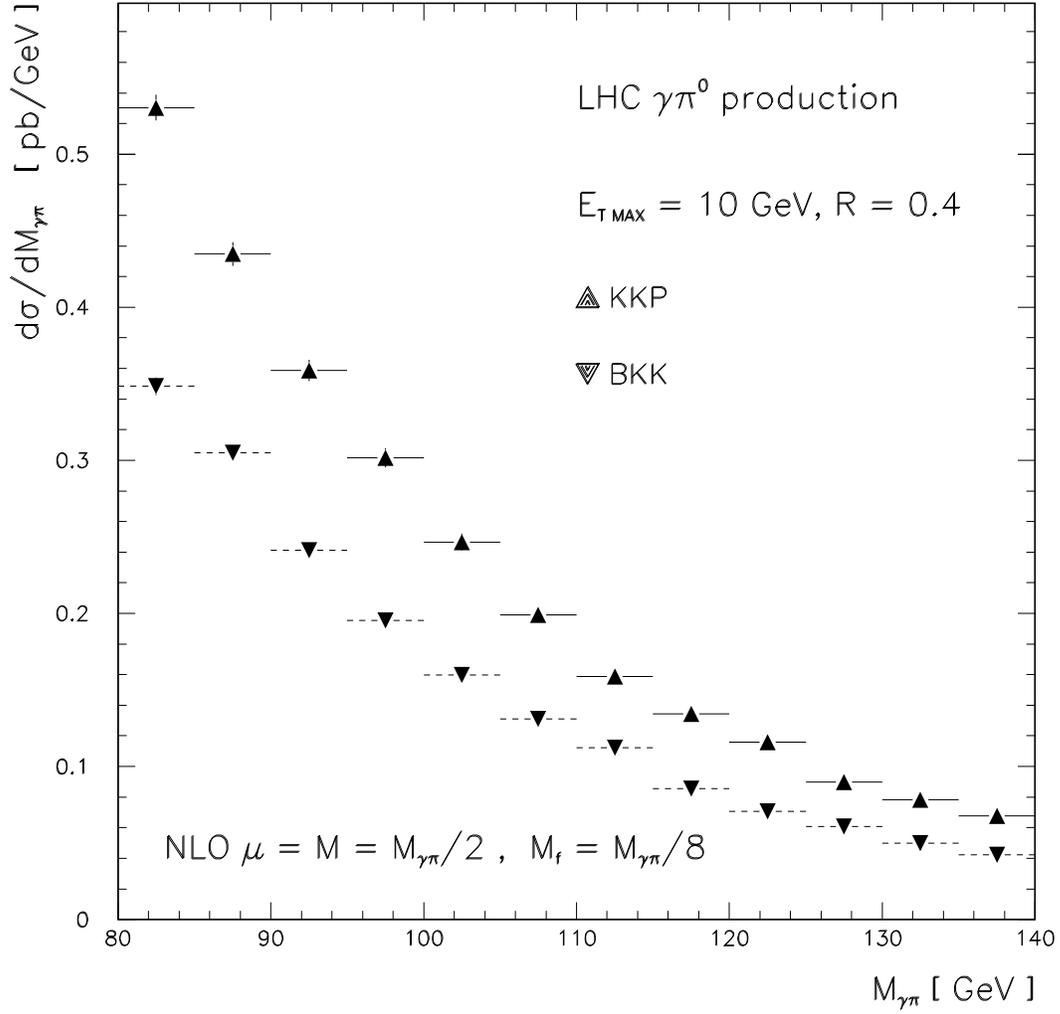}
\end{center}
\caption{\label{Fig:lhc_pg_mpg_kkp_bkk}{\em Comparison of the KKP and BKK fragmentation functions for $\pi^0 \, \gamma$ production}.}
\end{figure}

\begin{figure}[p]
\begin{center}
\includegraphics[width=\linewidth]{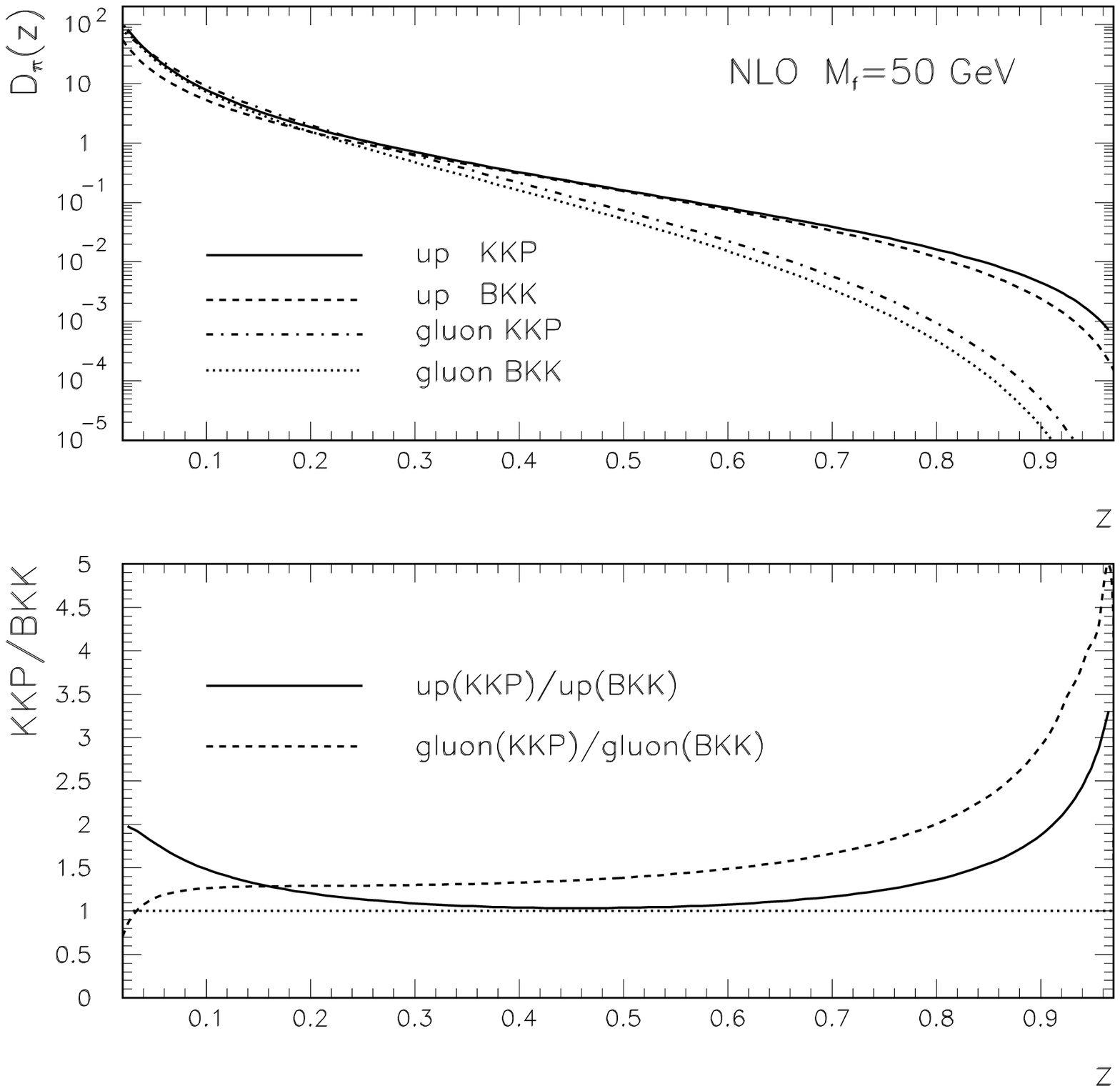}
\end{center}
\caption{\label{Fig:frag_pi_50}{\em Comparison of KKP and BKK fragmentation functions  for the fragmentation scale $M_f = 50$ GeV}}
\end{figure}

\noindent
%
Hard  isolation cuts test fragmentation functions
at high $z$ values, where current parametrisations are not at all
constrained by experimental data. 
This problem is visualised in 
Fig.~\ref{Fig:lhc_pg_mpg_kkp_bkk} where we show the photon--pion 
invariant mass distribution calculated with both, the KKP \cite{kkp} and the older BKK \cite{bkk} fragmentation functions. 
The KKP parametrisation leads to a higher 
prediction than the BKK one. The ratio of the two curves
(KKP/BKK) is about 3/2. For dipion observables the situation is much worse.
In Fig.~\ref{Fig:frag_pi_50} the discrepancy 
between the BKK and KKP parametrisations 
in the critical region is shown.
One observes that at high $z$ values the fragmentation 
gluon--to--pion is strongly
suppressed. Focusing on the quark--to--pion fragmentation, 
one sees that for high $z$ values the parametrisations differ
considerably. This will be discussed in more detail below.

\vspace{2mm}

\begin{figure}[p]
\begin{center}
\includegraphics[width=\linewidth]{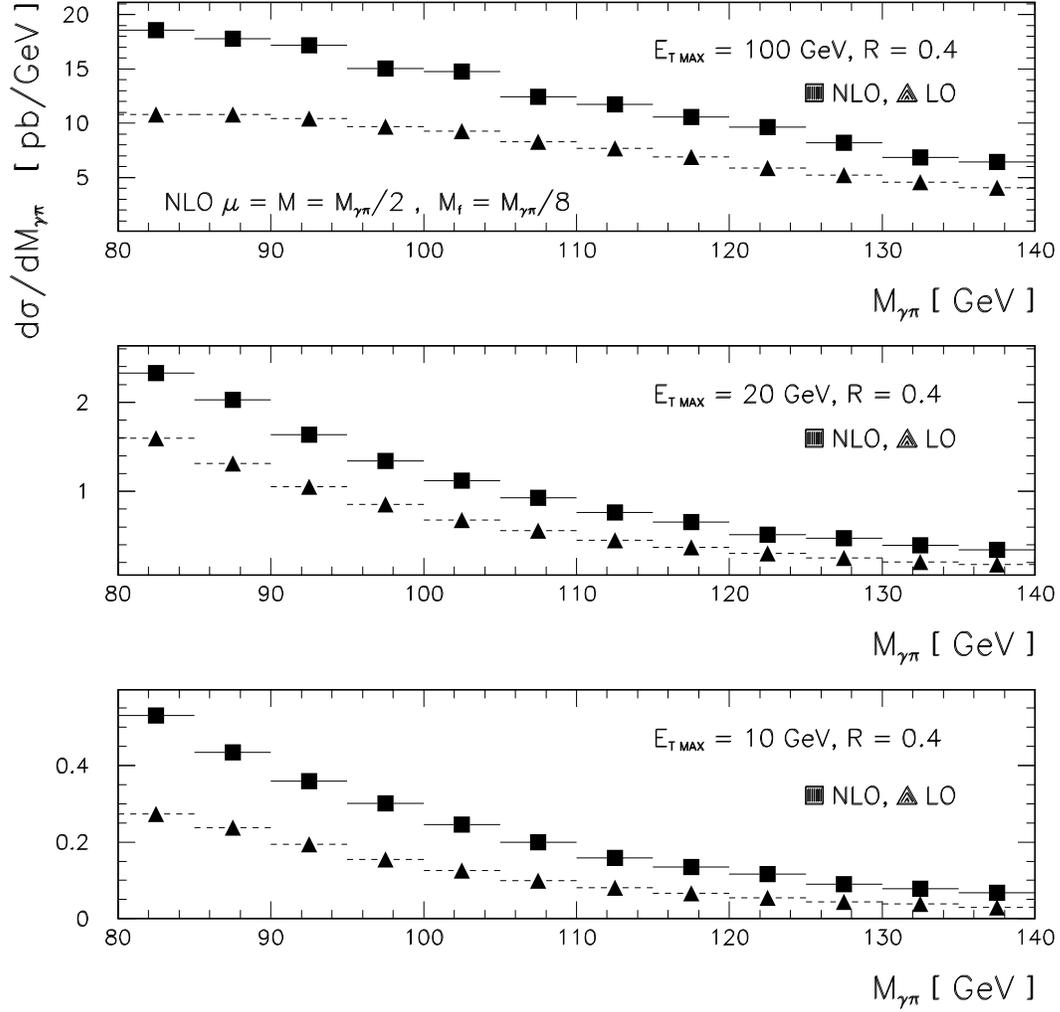}
\end{center}
\caption{\label{Fig:lhc_pg_mpg_nlo_lo}{\em Comparison of 
LO and NLO predictions of the invariant mass distribution of 
photon--pion pairs for various isolation criteria.} }
\end{figure}

\noindent
We have to comment now on the
remaining theoretical uncertainty due to higher
order corrections. As will be pointed out below the large $z$ region
is dangerous for the predictive power of NLO
calculations because terms
are produced which are divergent as $z\rightarrow 1$.
In Fig.~\ref{Fig:lhc_pg_mpg_nlo_lo} , we plot 
LO and NLO predictions\footnote{LO refers here 
to the Born matrix elements whereas NLO parton densities 
and fragmentation functions are used.} for the invariant mass distribution of the 
photon--pion pair for several isolation criteria.
Note that the ratio of both curves 
is pretty stable with decreasing $E_{T\;max}$. 
The ratio of both curves lies  
around 1.7, a typical value for a NLO correction. 
We just mention that the situation worsens
as $E_{T\;max}$ is reduced further
which indicates the presence of large NLO contributions 
induced by pressing $z$ to one. 
As will be pointed out below there are two reasons for that.
Phase space regions which are excluded at LO are populated at NLO
and become dominant, and large logarithms of $(1-z)$
are introduced.
To have a proper control
on the latter fixed order effect, a resummation of the relevant 
contributions is needed. For the  isolation
criteria used in the given plots we conclude 
from the NLO to LO ratios that these effects are not yet
dominant. Still, for harder isolation criteria the predictivity
of our fixed order approach has to be critically investigated.

\vspace{2mm}

\begin{figure}[p]
\begin{center}
\includegraphics[width=\linewidth]{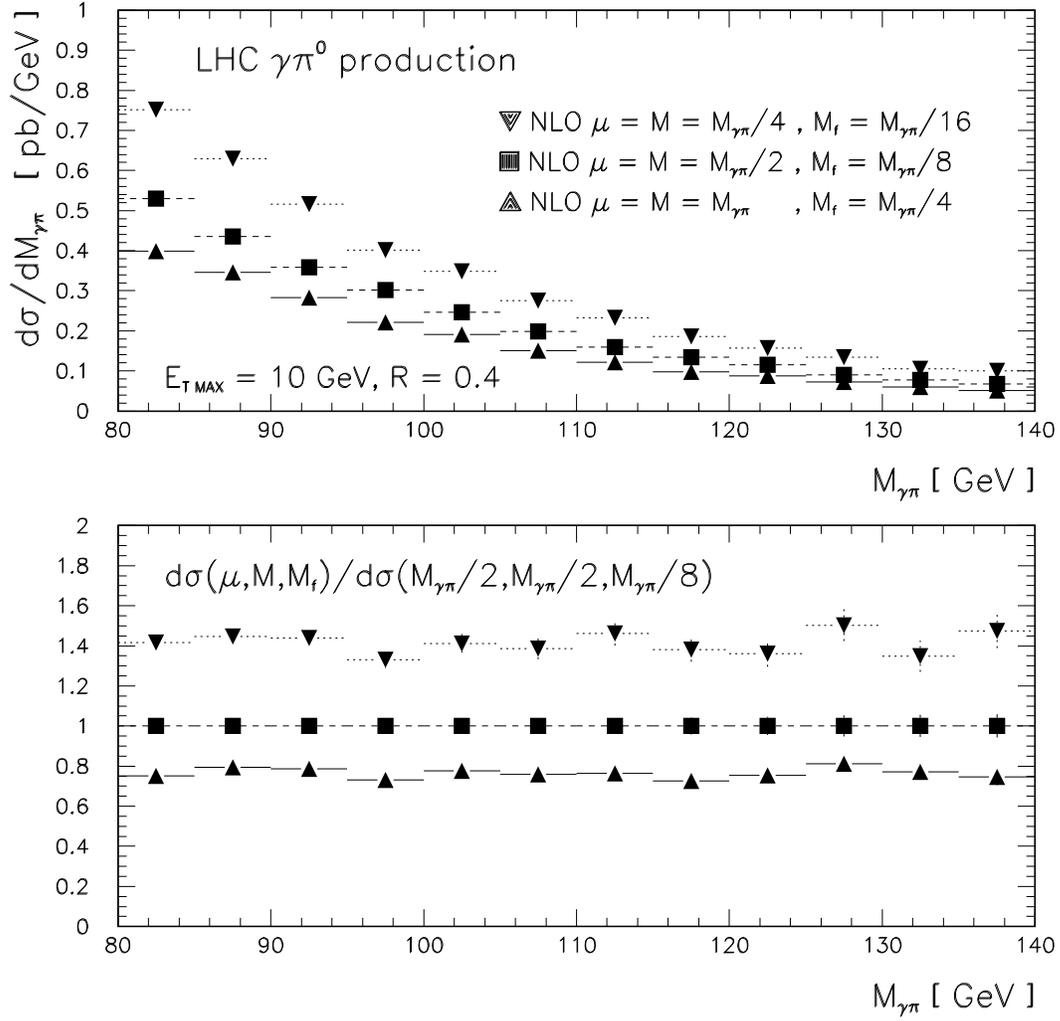}
\end{center}
\caption{\label{Fig:lhc_pg_mpg_scaledep}{\em Scale dependence of the 
incvariant mass distribution  of photon--pion pairs for the 
isolation criterion $E_{T max}=10$ GeV in the cone $R=0.4$.
The scales are varied around our standard choice $\mu=M=M_{\pi\gamma}/2$,
$M_f=M_{\pi\gamma}/8$ (boxes), as explained in the text, 
by multiplying all scales by 1/2 (inverse triangles) and 2 
(triangles) respectively.} }
\end{figure}

\noindent
As the isolation cuts spoil compensations between
scale dependent LO and NLO contributions the scale
dependence remains important. To give an idea of the 
uncertainties we have plotted the variation  
of the invariant mass spectrum of $\pi^0\gamma$ pairs
under the change of renormalization ($\mu$), factorisation ($M$)
and fragmentation  scale ($M_f$) for 
a strict isolation criterion in Fig.~\ref{Fig:lhc_pg_mpg_scaledep}.  
By multiplying our standard scale choice by 1/2 and 2
one finds variation of about +40 per cent to --30 per cent.
This variation is reduced if one uses less restrictive
experimental cuts. 

\vspace{2mm}

\begin{figure}[p]
\begin{center}
\includegraphics[width=\linewidth]{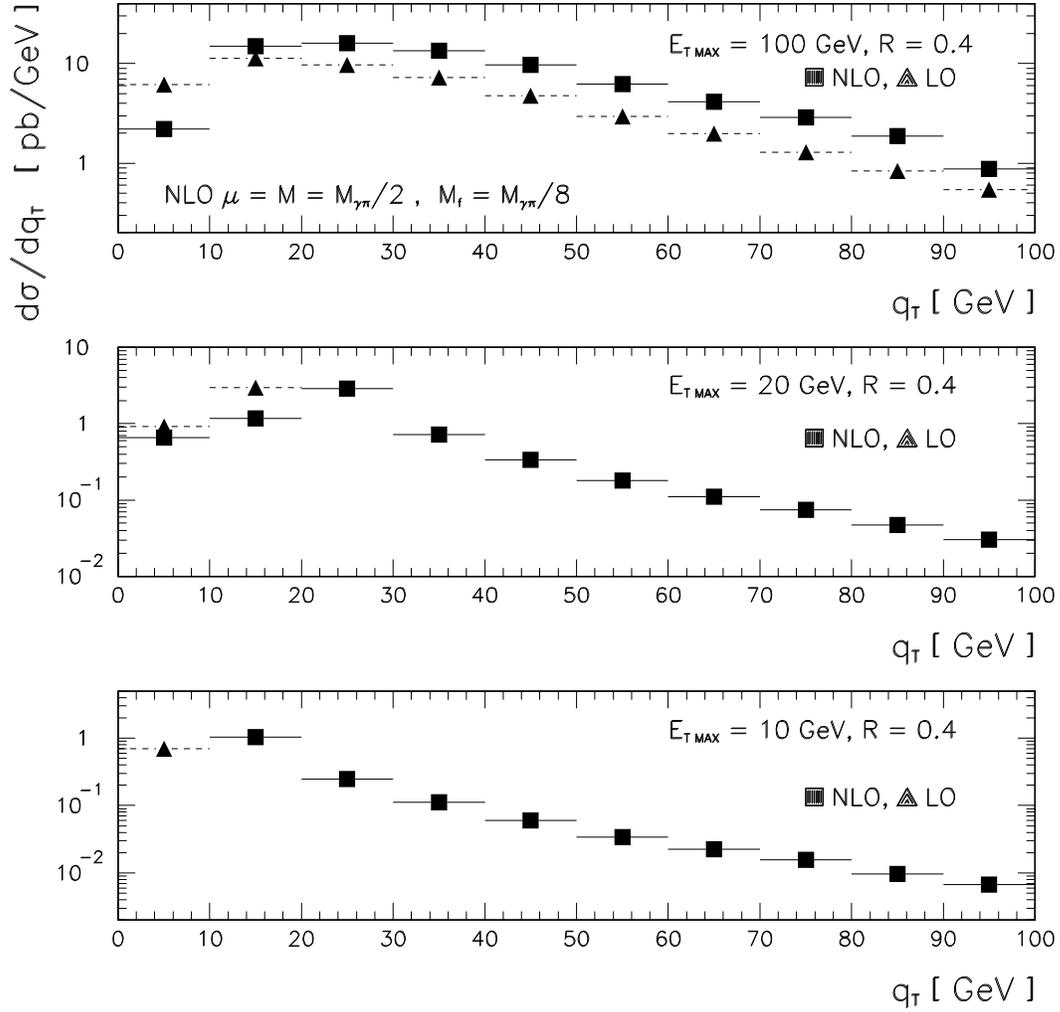}
\end{center}
\caption{\label{Fig:lhc_pg_qt_nlo_lo}{\em LO and NLO contributions 
for the $q_T$ distribution in $\pi^0\gamma$ production for 
different isolation criteria. The LO and NLO contributions 
populate different phase space regions when severe isolation is applied.} }
\end{figure}

\noindent
The importance of NLO calculations is best visible in distributions
which are restricted by LO kinematics. No LO parton shower
can provide for a correct description of events far away from
the LO kinematics. A relevant observable to exemplify this
point is the $q_T$ distribution of a photon--pion pair. 
Increasing the isolation criterion, suppresses the
LO phase space more and more, such that finally the whole
tail of the distribution is a pure NLO effect, see Fig.~\ref{Fig:lhc_pg_qt_nlo_lo}.
For completeness we show the effect of isolation
on this observable in Fig.~\ref{Fig:lhc_pg_qt_isol}.
For a comparison of the shape of our {\it DIPHOX} 
prediction to  PYTHIA see \cite{thomas-Kati}.

\section{Experimental cuts and 
higher order corrections}\label{isolpions} 

To understand the reliability of the presented predictions 
it is necessary to discuss the interplay between experimental cuts and
theoretical NLO results. QCD cross sections are 
stabilised by the inclusion of higher order corrections,
as scale dependencies are reduced. This works generally
the better the more inclusive the observables are.
Yet, the collider experiments at the Tevatron and the forthcoming LHC do not 
perform inclusive photon measurements. 
The experimental selection of
prompt photons requires isolation cuts.
We always use the cone isolation criterion 
introduced in Eq.~(\ref{isol}).
For the hadronic energy propagating into the isolation cone 
two sources can be distinguished in an experimental setup. 
First there is a component which stems from the
actual partonic, hard reaction which may be mapped to certain
Feynman diagrams -- this component is calculable inside
a QCD calculation to a given order. And second there is a
component coming from multiple-interactions and other events
in the same particle bunch (pile--up).
One has to be aware of the fact that an experimentally used
$E_{T \, max}$ is thus always bigger than a partonic
value used in our study. It is even so that
the experimental  values $E_{T \, max}$ are so small that a large fraction
of  the allowed hadronic transverse energy accompanying the candidate photon 
may come from such events. This leaves 
not much phase space for partonic fragmentation reactions.
In the following we mean with $E_{T \, max}$ only the
hadronic energy stemming from the hard interaction.
The sensitivity on $E_{T \, max}$ defined in Eq.~(\ref{isol}) 
is best seen  by looking at the fragmentation variable $z$,
which is the momentum fraction  of a hadron, 
$p_{T\,h}$ ($h=\pi,\gamma$), 
and its parent parton $p_T$.
At LO the kinematical constraints on this variable are induced 
by the experimental cuts through the collinearity assumption
in the fragmentation model, $p_T=z p_T + (1-z) p_T =p_{T\, h}+E_{T}^{had}$. 
One finds
\begin{eqnarray}
z > z_{min} = \frac{p_{T\,min}}{p_{T\,min}+E_{T\,max}}
\end{eqnarray} 
where $p_{T\,min}$ is the experimental cut on $p_{T\,h}$.
Evidently, the variable $z$ is pressed towards one, as 
the ratio $E_{T\,max}/p_{T\,min}$  decreases, means 
for hard isolation. At NLO, where extra radiated partons
are present, this bound is even more severe as 
the veto is more easily fulfilled.
The bulk of the pions which are produced in collider events 
correspond to values of $z$ substantially smaller than 1. 
Hard isolation tests only the very end of the $z$ spectrum.
The question is, wether we can 
reliably estimate the production rate of these high--$z$ pions, 
and especially of hard $\gamma \pi^{0}$ and $\pi^{0} \pi^{0}$ pairs.
We find three aspects important in this respect,
which are discussed in more detail in the following. 

\subsection{Reliability of fragmentation functions}\label{frag}

The usual fixed order perturbative QCD calculations of hadro
production -- as e.g. the one coded in {\it DIPHOX} -- rely on a model of
independent collinear fragmentation through fragmentation functions of a parton
of species $j$ (quark of any flavour, or gluon) into a pion,  
$D_{j \to \pi^{0}}(z,M^{2}_{f})$, $z$ is defined above. 
Let us first comment on the uncertainties at small $z$, where
this treatment is questionable. In particular 
the fixed order calculations do not resum the $\ln{(1/z)}$ arising order by 
order in perturbation theory, which become large in the small $z$ region. The
latter are associated with coherence effects in the multiple emission of soft
gluons, which control the inclusive production of small $z$ pions 
\cite{bcm,dkmt}. Therefore this usual framework is notoriously inadapted to
account for the production of these small $z$ pions. 
Moreover the dominant $z$ values at colliders are smaller than for fixed
targets, as a consequence of scaling violations. Therefore the erroneous
treatment of the fragmentation process at small $z$ will plague the estimates 
of inclusive dipion observables at colliders.
This is the reason why the fits of independent fragmentation functions 
$D_{i \to \pi^{0}}(z,M^{2}_{f})$ to experimental data presented in the 
literature \cite{bkk,kkp}, and which we use in this work, start above some 
minimum value for $z$, e.g. $z_{min} \sim 0.1$.
The independent fragmentation framework is expected to 
account correctly for the production of pions with larger $z$ values. 
The requirement of a loose isolation criterion such as, 
for example $E_{T \, max}~\leq$~100~GeV  for pions with 
$p_{T \, \pi^{0}}~\geq$~25~GeV imposes that 
$z  \geq 0.2$ over the 
whole $p_{T \, \pi^{0}}$ range of interest. 
To avoid uncertainties due to small $z$ values, as was pointed out 
already, we do not provide estimates for 
inclusive observables, but only for observables corresponding to pions 
submitted to isolation cuts. 
\vspace{0.2cm}

\noindent
The behaviour near the end point, $z \rightarrow 1$,  is
responsible for most of the uncertainties that plague the present study.
Indeed, the use of very tight isolation cuts such as those used by the collider
experiments select the vicinity of the elastic boundary $z \sim 1$.       
For example, in the Run Ib of the
Tevatron, for prompt photon measurements, CDF required $E_{T \, max} = 2$ GeV
in a cone $R = 0.7$. For $p_{T} \geq 15$ GeV, this means $z_{min}(p_{T}) \geq
 0.9$. The value $E_{T \, max} = 2$ GeV required by the CDF experiment includes also the deposition of hadronic transverse energy coming from underlying events, while the latter are ignored in our partonic calculation as noted above. 
Consequently, the actual effective value of $E_{T \, max}$ which should be used in a
partonic calculation should be even smaller, the effective $z_{min}$ would be even larger.
However the fragmentation functions 
$D_{j \to\pi^{0}}(z,M^{2}_{f})$ are fitted to data corresponding essentially
to  $z \leq 0.7$, i.e. much below the relevant region for the present study.
Moreover, the data used in \cite{bkk,kkp} concern hadronic final states in
$e^{+}e^{-}$ annihilation. The gluon--to--pion fragmentation function is poorly
constrained at moderate and large $z$ by this process because the subprocesses
involving gluons in $e^{+}e^{-}$ annihilation appear at higher order in the
perturbative expansion in powers of $\alpha_{s}$, while subprocesses involving
outgoing gluons contribute at lowest order in hadronic collisions. Although the
contributions from gluon fragmentation are not expected to dominate, large
uncertainties on their magnitude may still affect quantitative predictions.
In order to illustrate how poor the present-day control over the large $z$
region is, it is instructive to compare the extrapolations of the KKP and BKK sets of parton-to-$\pi$ fragmentation functions. The KKP set is the most recent one, and it is constrained by more data, especially the LEP1 data at the $Z$ peak, in the region $0.1 \leq z \leq 0.7$. These two sets use different analytic ans\"atze. These different ans\"atze do not differ very much in the $z$ range 
in which they are most constrained by the $e^{+}e^{-}$ data of course, but 
their respective extrapolations turn out to differ substantially from each 
other above $z \geq 0.7$, especially between $z = 0.9$ and 1, as shown 
in Fig.~\ref{Fig:frag_pi_50}. 
This is true for the gluon, but also, though to a lesser extend, for 
the $u$ and $d$ fragmentation functions. This translates into a large 
uncertainty for the predictions of the reducible Higgs boson background 
in the channel $H \to \gamma \gamma$ in the mass range 80--140 GeV. 
Clearly a determination of 
fragmentation above $z \geq 0.7$ from data sensitive to this range is required 
for any further accurate prediction.
Such an experimental input could in principle
be provided 
by data on inclusive $\pi^0$ production at fixed target energy 
\cite{bgpw3}\footnote{In Ref.~\cite{afgkw}, the comparison between E706
data and BKK fragmentation functions reveals an overall normalisation discrepancy. However, as noticed in Ref.~\cite{bgpw3}, this problem seems 
to be largely reduced by using the KKP set.}, 
as well as measurements of isolated pions by the collider experiments CDF
and D0, in association with their prompt photon measurements. 
\vspace{0.2cm}

\noindent
It is worth mentioning that the present inaccuracy of the description of
fragmentation at large $z$ is not only a problem confined to the type of 
perturbative QCD calculations which are discussed in this article. They also
potentially affect the fragmentation stage of the ``full hadronic event
generators" such as PYTHIA \cite{pythia} and HERWIG \cite{herwig} as well.
Indeed, the respective fragmentation ingredients of these event generators have
been tuned to describe accurately the bulk of the data, especially at LEP, 
whereas the tails of the distributions near the end point have,
to our knowledge, never been analysed in detail. In addition, the parton
showers implemented in these event generators provide an effective summation of
the soft gluon effects near $z \sim 1$ at a leading logarithmic accuracy; on
the other hand they miss a part of the higher order corrections to the partonic
hard subprocesses which may be large. The respective approximations in the
``partonic calculations" on one hand, the full event generators on the other
hand, being different, it would be instructive to compare the results given by
both approaches\cite{thomas-Kati}. 

\subsection{Large $z$ values and the need for end-point summation}

As the production of isolated pions at colliders and the one of inclusive
pions at fixed target experiment, deal with $z$ values near the end point  
$z = 1$, the theoretical calculation of the partonic subprocesses is confronted
with the appearance of powers of large logarithmic terms $\ln{(1-z)}$ to any
fixed order in perturbation theory. These large logarithms have in principle to
be summed to all orders in order to improve -- if not restore -- the predictive
power of the partonic calculation. 
The general framework to resum infrared logarithms in hadronic collisions
has been settled in \cite{catani-colour,sterman-colour}, 
however further work is needed to treat the endpoint summation in the case of one particle production,
even more for pair production, where the kinematics is more constrained.   
In the present study, we investigate what can be done without such summations, relying
on the extrapolations of the fragmentation functions available in the
literature, and examining critically the limitations of our approach. 

\vspace{0.2cm}

\noindent
To be explicit let us consider the process $PP\rightarrow \gamma \pi^0 + X$
in detail. The photon can be either direct or from fragmentation.
The second case is considerably suppressed if one imposes severe
isolation; thus we deal with the direct process only.
At leading order there are two types of  partonic processes:
$q g\rightarrow q \gamma$\footnote{Here $q$ stands for quark or antiquark.}, $q\bar q\rightarrow \gamma g$. At small $x$ values typical for the LHC the 
first is dominant.
Radiative corrections give rise to initial state
and final state singularities. Only the latter are of relevance
for the point to be made here. They arise from the 
splitting of the outgoing parent quark into a quark and a gluon. Both
may fragment into a pion but, again, for severe isolation
one may focus on the dominant $q\rightarrow \pi^0$ fragmentation only.
Thus we are looking at the case where a collinear gluon is radiated of
a hard quark, indicated by the subscript $g//q$, 
which subsequently fragments into a pion.
Following Ref.~\cite{bgpw} the finite remnant of the collinear 
approximation of this correction is of the form
\begin{eqnarray}
\sigma(PP\rightarrow qg\gamma)|_{g//q} &=&
\int dy_\pi dy_\gamma dp_{T\,\pi}
\int\limits_{x_{min}}^{1} \frac{dx}{x}
\int\limits_{z_{min}}^{1} \frac{dz}{z}
\frac{\alpha_s(\mu)}{2\pi}
\nonumber \\ && \times
D_{\pi^0/q}(x,M_f^2) \frac{F_{q/P}(x_1,M^2)}{x_1}
\frac{F_{g/P}(x_2,M^2)}{x_2} | {\cal M}_{qg\rightarrow q\gamma}^{LO} |
\nonumber \\ && \times
\left[ 2 \left( \frac{\ln(1-z)}{1-z} \right)_+ a_{qq}^{(4)}(z)
+ \ln\left( \frac{p_{T\,\pi}^2}{x^2z^2M_f^2}\right) P^{(4)}_{qq}(z)
+ \dots \right] \\
P^{(4)}_{qq}(z) &=& \frac{a_{qq}^{(4)}(z)}{(1-z)_+} + \frac{3}{2}C_F \delta(1-z) \nonumber\\
a_{qq}^{(4)}(z) &=& C_F (1+z^2) \nonumber
\end{eqnarray}
Here $y_\pi,p_{T\,\pi}$ are the rapidity and transverse momentum of the pion,
$y_\gamma$ the rapidity of the photon, $xz$ the momentum fraction
of the pion with respect to the parent quark, $p_{T\,\pi}=xzp_{T\,q}$, 
and $z$ the momentum fraction
of the gluon with respect to the parent quark, $p_{T\,g}=z p_{T\,q}$.
In the given collinear approximation the momentum fractions
of the initial partons are 
\begin{eqnarray}
x_1&=& \frac{p_{T\;q}+p_{T\;g}}{\sqrt{S}}[\exp(y_\pi)+\exp(y_\gamma)]\nonumber\\
x_2&=& \frac{p_{T\;q}+p_{T\;g}}{\sqrt{S}}
[\exp(-y_\pi)+\exp(-y_\gamma)]\nonumber\\
z_{min} &=& x_{min}/x\nonumber
\end{eqnarray}
The value of $x_{min}$ will be discussed
below, as it is the sensitive variable to isolation cuts. 
The dots stand for terms which are less dominant as $z\rightarrow 1$.
They are not necessarily small.  
By writing $a_{qq}^{(4)}(z)/z =
C_F[(1-z)^2/z+2]$, one sees that the first term in the bracket only leads to
non logarithmic corrections.  Thus, approximating the bracket by 2
and using that for $z\rightarrow 1$, $\ln(z)={\cal O}(1-z)$,
the $z$ integration can be done explicitly and one finds
\begin{eqnarray}
\sigma(PP\rightarrow qg\gamma)|_{g//q} &=&
\int dy_\pi dy_\gamma dp_{T\,\pi}
\int\limits_{x_{min}}^{1} \frac{dx}{x}
\frac{\alpha_s(\mu)C_F}{\pi}
\nonumber \\ && \times
D_{\pi^0/q}(x,M_f^2) \frac{F_{q/P}(x_1,M^2)}{x_1}
\frac{F_{g/P}(x_2,M^2)}{x_2} | {\cal M}_{qg\rightarrow q\gamma}^{LO} |
\nonumber \\ && \times
\left[ \ln(1-x_{min}/x)^2  
+ \ln\left( \frac{p_{T\,\pi}^2}{x^2 M_f^2}\right) \ln(1-x_{min}/x) 
+ \dots \right]
\end{eqnarray} 
The important point is that 
the behaviour of these terms changes drastically as one goes from the inclusive
case, where $x>x_{min}=2p_{T\,\pi}/\sqrt{S}$ is much smaller than one, 
to the case of severe isolation which press $x$ to  values near one. 
Using the fact that the fragmentation function
behaves as $D_{\pi^0/q}(x) \sim N (1-x)^\beta$ with $\beta\sim 1$,
one can write 
\begin{eqnarray}
\frac{D_{\pi^0/q}(x)}{x} = \frac{D_{\pi^0/q}(x_{min})}{x_{min}}
\left( 1-\frac{x-x_{min}}{1-x_{min}} \right) + {\cal O}(1-x_{min},1-\beta)
\end{eqnarray}
and perform the integral over $x$. Focusing on 
the double logarithmic terms, one finds
\begin{eqnarray}
\sigma(PP\rightarrow qg\gamma)|_{g//q} &=&
\int dy_\pi dy_\gamma dp_{T\,\pi}
\frac{\alpha_s(\mu)C_F}{\pi}
\nonumber \\ && \times (1-x_{min})
\frac{D_{\pi^0/q}(x_{min},M_f^2)}{x_{min}} \frac{F_{q/P}(x_1,M^2)}{x_1}
\frac{F_{g/P}(x_2,M^2)}{x_2} | {\cal M}_{qg\rightarrow q\gamma}^{LO} |
\nonumber \\ && \times
\frac{1}{2}\left[ \ln(1-x_{min})^2  
+ \ln\left( \frac{p_{T\,\pi}^2}{M_f^2}\right) \ln(1-x_{min}) 
+ \dots \right]
\end{eqnarray} 
the dots stand for single logarithmic terms and terms which are
suppressed by an extra factor of $(1-x_{min})$ and $(1-\beta)$.
Note that outside the bracket, up to the factor
$\alpha_s(\mu)C_F/(2\pi)$, one has just the approximation
for the leading order contribution. This means that the bracket
itself times this factor could be taken as a good approximation 
for the higher order part of a $K$ 
factor\footnote{Note the way this $K$ factor depends 
on the experimental cuts applied and on the scale choice. 
This is a nice example of the fact
that for non inclusive quantities the  naive use of $K$ factors
is not at all adequate to take into account higher order corrections.}.     
\begin{eqnarray}
K = 1 + \frac{\alpha_s(\mu)}{\pi}\left[ \ln(1-x_{min})^2  
+ \ln\left( \frac{p_{T\,\pi}^2}{M_f^2}\right) \ln(1-x_{min}) 
+ \dots \right]
\end{eqnarray}
In here a factor of two is included from the  second process with initial partons $\bar q g$.
To get a feeling for the typical scales in the case of isolation
one may try to choose $M_f$ such that the large logarithms compensate
for a $p_{T\,\pi} \sim p_{T\,min}$.
This can be achieved by a (optimal) scale choice
\begin{eqnarray}\label{opti-scale} 
M_f^{opt} = p_{T\pi\,min} \sqrt{\frac{E_{T\,max}}{p_{T\,min}+E_{T\,max}}}
\end{eqnarray}  
Plugging in $p_{T\,min}=25$ GeV \footnote{Due to the NLO kinematics
$p_{T\,\gamma} \ge p_{T\,\pi}$. For asymmetric $p_T$ cuts the photon
will always be the particle restricted by the larger $p_T$ cut.} and 
$E_{T\,max}=10$ GeV one finds $M_f^{opt}\sim 10$ GeV. 
This motivated us to choose for $M_f$ in the figures presented  not
the canonical choice $M_{\pi\gamma}/2$ 
but rather $M_{\pi\gamma}/8$. Note that for a much higher
scale choice  the logarithms become constructive. 
For the upper choice of parameters 
and using $M_f=2 p_{T\,min}$ arbitrarily 
large K factors are induced in the limit $x_{min}\rightarrow 1$.   
The fact that the optimal
scale in Eq.~(\ref{opti-scale}) goes to zero, as $E_{T\,max}$
goes to zero indicates that the cone isolation criterion
is not infrared safe in this limit. In other words, if the experimental
value of  $E_{T\,max}$  is already saturated by 
events from the hadronic environment such that 
there is no phase space left for the hard parton, 
fixed order calculations are not applicable. 

\vspace{0.2cm}
\noindent
It should be stressed that the whole discussion 
was neglecting constant terms and sub-leading logarithms.
Thus it should only give
a rough idea of the much more involved,  complete situation.
Using the same approximations the  exercise can be worked through
keeping also non--leading terms.  
The result is that
one encounters in realistic situations  additive 
contributions which are of the same size as the considered ones. 
All this deserves further investigations especially in the
light of a possible resummation of the
large logarithms and the role of these terms.   

\subsection{Hard isolation and asymmetric $p_T$ cuts}

\begin{figure}[p]
\begin{center}
\includegraphics[width=\linewidth]{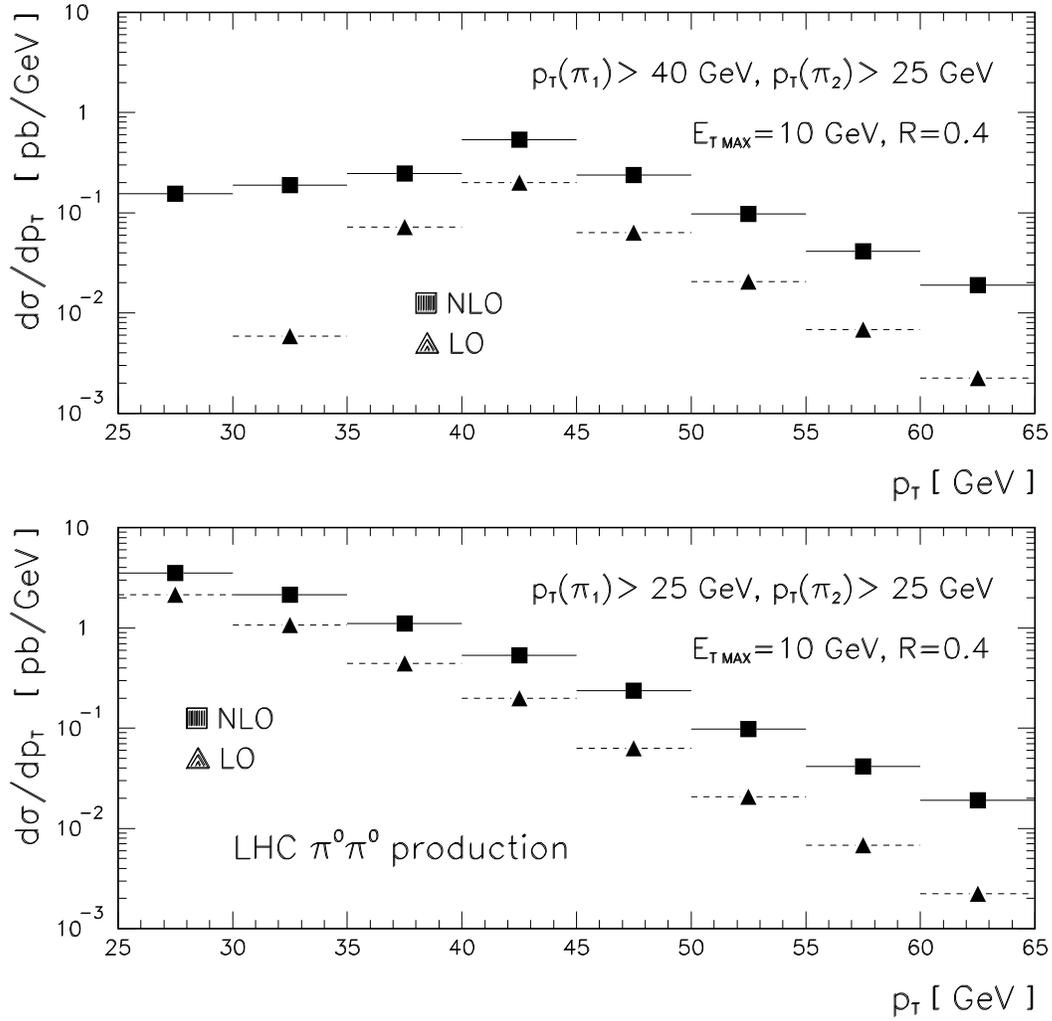}
\end{center}
\caption{\label{Fig:lhc_pp_pt_sym_asym}{\em The transverse momentum 
spectrum of the pions in dipion production 
for symmetric (bottom) and asymmetric (top) $p_T$ cuts. Scales: 
$\mu=M=M_{\pi\gamma}/2$, $M_f=M_{\pi\gamma}/8$.}}
\end{figure}

In LHC studies concerning Higgs boson search 
the used $p_T$ cuts are commonly asymmetric. 
We want to make the point here that this jeopardizes the 
predictive power of NLO calculations. The interplay between such asymmetric
cuts and hard isolation criteria leads at LO to vetoed 
phase space regions which are filled by NLO processes
only. 
The effect is visualised in Fig.~\ref{Fig:lhc_pp_pt_sym_asym}, where
the LO and NLO {\it DIPHOX} prediction for the dipion $p_T$ spectrum  
is plotted for symmetric, both $p_T>25$ GeV, and asymmetric cuts,
$p_{T 1}>40$ GeV, $p_{T 2}>25$ GeV.
In the case of asymmetric cuts one observes that the LO component is suppressed for low $p_T$ values. This can be understood as follows:\\
Due to transverse momentum conservation, one has
\begin{equation}\label{ptbalance}
p_{T\, 1} + E_{T\,1}^{had} = p_{T\, 2} + E_{T\,2}^{had}
\end{equation} 
where the pion 1 is assumed to have higher $p_T$ than pion 2.
This means for the transverse momentum of the pair, $q_T$ that
\begin{equation} \label{qtbound}
0\leq q_T = p_{T\, 1} - p_{T\, 2} = 
E_{T\,2}^{had} - E_{T\,1}^{had} \leq E_T^{max}
\end{equation}  
and because $p_{T\,1}>p_{T\,1\,min}$ (= 40 GeV in our case)
\begin{equation} \label{ptbound}
p_{T\, 2} \geq p_{T\, 1} - E_T^{max}  \geq p_{T\,1\,min} - E_T^{max} 
\end{equation}
This means that in the case of $p_{T\,1\,min} - E_T^{max}>p_{T\,2\,min}$
one has a phase space restriction coming from the isolation criterion.
Note that a saturation of the bound, 
$p_{T\, 2} = p_{T\,1\,min} - E_T^{max}$ 
is only possible if $E_{T\,1}^{had}\rightarrow 0$, i.e.
$z_1\rightarrow 1$. But in this limit the fragmentation
function goes to zero and suppresses the cross section
additionally. This will not happen in the case of
photon--pion production where the same formulas apply
with $E_{T\,1}^{had}=0$. Eq.~(\ref{ptbound}) is also
true and leads to a restriction of the phase space 
but there is no additional suppression. 
The problem aggravates for lower $E_{T max}$ values but becomes
less relevant when $E_{T max}>p_{T min 1}-p_{T min 2}$
and vanishes in the case of symmetric cuts $p_{T min 1}=p_{T min 2}$.
The hole in the $p_T$ spectrum will be filled 
with events from higher order corrections.
As the low $p_T$ region is energetically preferred, one gets
numerically important NLO contribution to the cross section
in the dipion case.  
The ratio between LO and NLO can become very big due to this effect.
As pointed out, in photon--pion production the LO order contribution is less
suppressed and the effect is milder.
In the $q_T$ distribution shown in Fig.~\ref{Fig:lhc_pg_qt_isol},
one observes that only the first bin contains the LO contribution
due to the same kinematical restriction mentioned above.
On the other hand the NLO contribution is basically zero
there (even below the plotted range). This is because 
pions with $p_T$'s as large
as the photon $p_T$ are disfavoured. 
The steep fall of the quark--to--pion fragmentation functions
makes it very unlikely to find a photon--pion pair with 
$p_{T \gamma}\sim p_{T \pi}>40$ GeV which is needed to have $q_T\sim 0$. 

\vspace{0.2cm}
\noindent
From the discussion one sees that symmetric cuts are preferable 
from theoretical side
for processes which have to be suppressed by isolation.
Whereas this effect is for sure important for dipion
production the effect is numerically sizable in the
photon--pion case only for isolation criteria which are harder 
than the ones given in the presented figures.
As can be inferred from Fig. \ref{Fig:lhc_pg_mpg_nlo_lo}, 
the ratio of the NLO to LO
curves does not explode because the discussed phase space effect
is not yet numerically sizable.  Still we want to make the
point that our NLO analysis suggest to choose symmetric
transverse momentum cuts as theoretical higher order
predictions tend to be more reliable in that case.

\section{Conclusions and outlook}\label{final}

We have presented a full next to leading order study of the production of 
$\pi^{0}\gamma$ as well as $\pi^{0}$ pairs with a large invariant
mass relying on the computer code {\it DIPHOX}. 
We have presented estimates for invariant mass 
and transverse momentum of the pair ($q_T$) distributions
taking into account realistic experimental cuts.  
The NLO corrections are large.  
Theoretical calculation improves the scale dependences in the
case of inclusive observables.
On the other hand, the results remain plagued by large uncertainties even in
the NLO approximation, if realistic isolation criteria
are applied. One source of large uncertainties is our poor present
knowledge of the parton--to--pion fragmentation functions at large $z$. The
estimates which we have presented here rely on extrapolations of analytical
ans\"atze for these fragmentation functions for values of $z$ above 0.8, i.e.
outside the region  where they were fitted to the data. The spreading of the
extrapolations of  different sets of parametrisations available in the
literature for $z$ above 0.8 illustrates this fact.  
Experimental information may be provided by inclusive
measurements of single pions and pions pairs at fixed targets, e.g. as
performed by the E706 experiment \cite{e706}, as well as by measurements of
isolated pions at colliders, as
could be performed by the CDF \cite{cdf} and D0 \cite{d0} experiments.

\vspace{0.2cm}

\noindent
Apart from this more experimental issue we pointed out that
our next-to-leading order approach is still plagued by relatively
large uncertainties due to phase space effects related to the 
different LO  and NLO kinematics and also the presence of
potentially large logarithms $\ln(1-x_{min})$, if severe
isolation cuts are imposed.
Concerning the phase space effect, we discussed that symmetric
$p_T$ cuts for the pions/photons are preferable  
from theoretical side. Otherwise NLO corrections are magnified
in phase space regions vetoed at lowest order.
Our predictions for $\gamma\pi^0$ production seems 
still  reliable, up to scale variations of around 40 per cent, 
for the used isolation cuts. 
The same cannot be said for $\pi^0\pi^0$ production.
The region 
$z\rightarrow 1$ is probed in a more critical way in this case. 
Further, predictions are very sensitive to the unconstrained part
of fragmentation functions.
We have to conclude that the uncertainties for this 
component of the reducible background for Higgs searches are not 
under control from theoretical side even at the 
next--to--leading order level. We only quote 
that we find large K factors and large 
scale variations, factors of 3 to 4 respectively, 
if we use the same isolation cuts as was done for the
isolated $\gamma\pi^0$ production case. 
Fortunately this component to the diphoton background
can be suppressed experimentally by separating pions from photons
by  distinguishing the shapes of the corresponding
electromagnetic showers. Reduction factors of 2 to 3 per 
pion are possible and compensate somewhat the poor knowledge 
of the pion rates. Still, a more satisfactory theoretical 
knowledge of the dipion and pion/photon rates
at LHC requires a better understanding of fragmentation functions
near $z=1$, the resummation of 
large endpoint logarithms of $(1-z)$, together with the use of experimental
cuts which do not destabilise compensations between 
LO and higher order contributions.

\vspace{0.2cm}

\begin{figure}[p]
\begin{center}
\includegraphics[width=\linewidth]{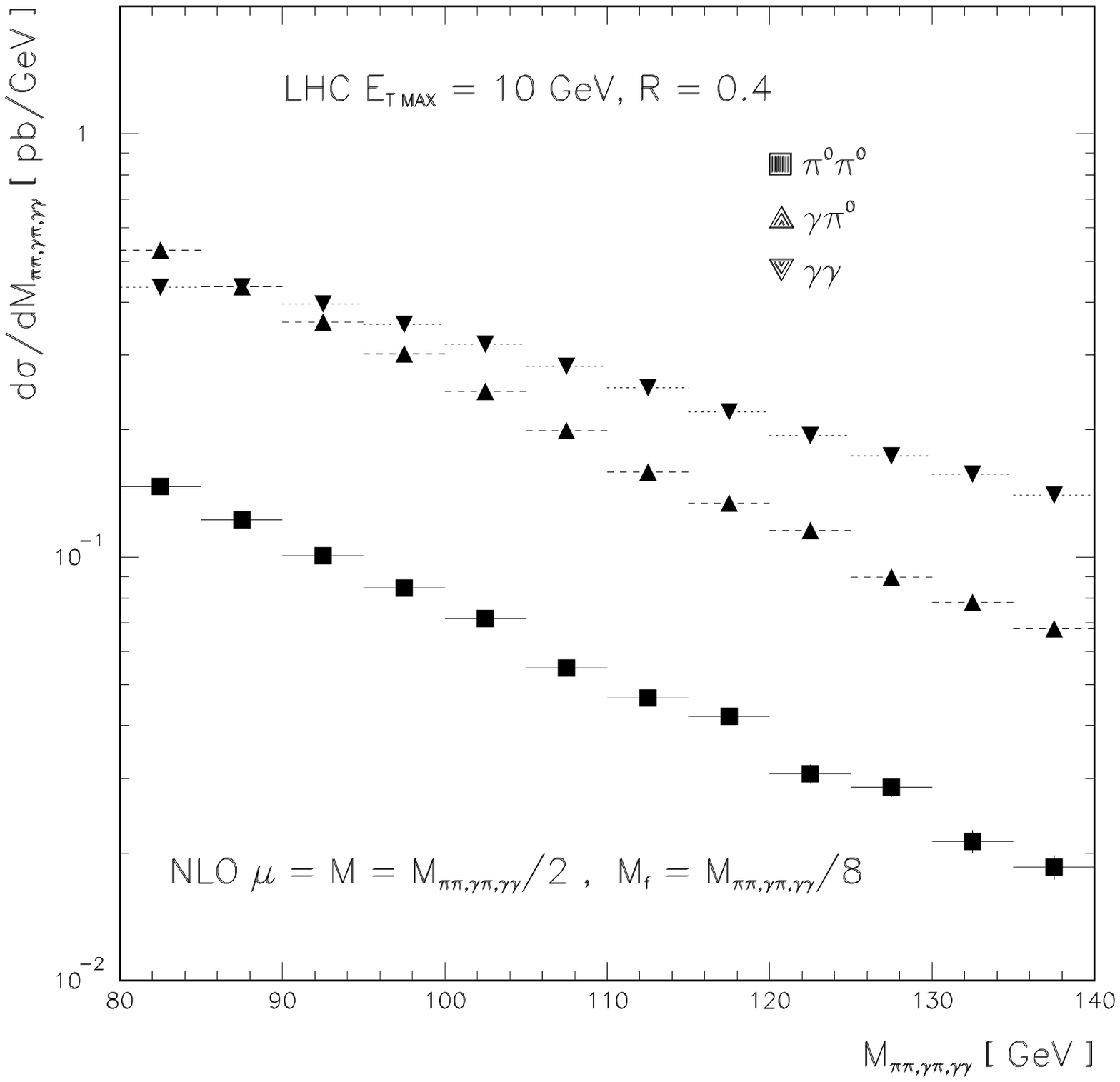}
\end{center}
\caption{\label{Fig:lhc_all_et10}{\em Comparison of the $\gamma\gamma$, $\gamma\pi^0$ and $\pi^0\pi^0$ invariant mass distribution for the 
isolation cuts $E_{T max} = 10$ GeV, R=0.4.}}
\end{figure}

\noindent  
From the discussion above it should be clear that it is a delicate issue 
to make definite predictions for the $\gamma\pi^0$ and $\pi^0\pi^0$  component
of the $H\rightarrow\gamma\gamma$ background.
All statements depend strongly on the experimental cuts. Especially severe
isolation cuts restrict the predictivity of our result. To make
some explicit statements we just want to remind how the  
$\gamma\pi^0$, $\pi^0\pi^0$ rates compare to the irreducible
$\gamma\gamma$ background for a definite isolation
criterion, see Fig.~\ref{Fig:lhc_all_et10}. 
As the Higgs boson of the Standard Model is already restricted by the LEP
experiments to have a mass of $M_H\ge 115$ GeV, the $\gamma\pi^0$ rate
is about a factor of two below the irreducible  $\gamma\gamma$ background.
Uncertainties due to fragmentation functions and scale variations
may be factors of around two. 
On the other hand additional experimental capabilities based 
e.g. on calorimetric studies 
indicate that the rejection factor of pions w.r.t 
isolated photons can be improved by at least a factor two \cite{atlas,cms}.
These numbers suggest a conservative upper bound on the ratio
of the $\gamma\pi^0$, $\gamma\gamma$ rates of $0.5$. If one tightens the isolation
criterion this will be further reduced, though one has to keep in mind that 
the theoretical uncertainty is getting more important. In this respect we note
that a comparison with PYTHIA distributions shows that
our rates are lower if compared to default PYTHIA settings \cite{thomas-Kati}. 
Though we want to stress the relevance of a next--to--leading order
analysis, from our discussion one may not expect that the  
$\gamma\pi^0$ process becomes a dangerous background. 
As already noted above the uncertainties from 
fragmentation functions and scale variation for
the $\pi^0\pi^0$ rate are substantial. To be conservative
factors as big as ten should be considered. Assuming again
a reduction factor of two per pion from pion--photon separation 
would lead in Fig.~\ref{Fig:lhc_all_et10} to a  
$\pi^0\pi^0$ to $\gamma\gamma$ ratio of around 1/3 for $M_{\pi\pi}>115$ GeV.
More severe isolation will suppress this ratio considerably 
again with the price of even increased uncertainties.
Note that the given numbers should only give an idea of the  situation,
as they depend strongly on the experimental settings.    
We can tentatively conclude that next--to--leading order corrections 
should not change the qualitative statement that 
the $\pi^0\pi^0$, $\pi^0\gamma$ backgrounds are sufficiently under 
control at the LHC, although a couple of interesting QCD issues 
plaguing them still deserve  to be addressed.

\vspace{0.2cm}

\noindent
\section*{Acknowledgements} We thank K. Lassila--Perini, V. Tisserand
and E. Tournefier for discussions and comments. 
This work was supported in part by the EU 
Fourth Training Programme ``Training and Mobility of Researchers", Network
``Quantum Chromodynamics and the Deep Structure of Elementary Particles",
contract FMRX-CT98-0194 (DG 12 - MIHT). LAPTH is a ``Unit\'e Mixte de
Recherche (UMR 5108) du CNRS associ\'ee \`a l'Universit\'e de Savoie".

\newpage

\end{document}